\documentstyle[12pt]{article}
\def\beq{\begin{equation}} 
\def\eeq{\end{equation}} 
\def\bea{\begin{eqnarray}}  
\def\eea{\end{eqnarray}}  
\def\bq{\begin{quote}}  
\def\eq{\end{quote}}  
  
\def\beqa{\begin{eqnarray}}  
\def\eeqa{\end{eqnarray}}  
\def\be{\begin{equation}}  
\def\ee{\end{equation}}  
\def\beq{\begin{equation}}  
\def\eeq{\end{equation}}

\def\pa{\partial}  
\def\kaps{{\kappa}^{2}}

\def\bi{\begin{itemize}}  
\def\ei{\end{itemize}}

\def\ov{\overline}  
\def\nn{\nonumber \\}

\def\lc{{\cal L}}  
\def\ca{{\cal A}}

\begin{document} 

\pagestyle{empty}
\begin{flushright}   hep-th/0004093
\end{flushright}
\vskip 2cm
\begin{center}
{ \Large  \bf Supersymmetrizing Branes with Bulk\\
in Five-Dimensional Supergravity}
\vspace*{5mm} \vspace*{1cm} 
\end{center}
\vspace*{5mm} \noindent
\vskip 0.5cm
\centerline{\large Adam Falkowski${}^1$ Zygmunt Lalak${}^{1,2}$
and Stefan Pokorski${}^{1}$}
\vskip 1cm
\centerline{\em ${}^{1}$Institute of Theoretical Physics}
\centerline{\em University of Warsaw, Poland}
\vskip 0.3cm
\centerline{\em ${}^{2}$Physikalisches Institut, Universit\"at Bonn}
\centerline{\em Nussallee 12, D-53115 Bonn, Germany}
\vskip2cm

\centerline{\bf Abstract}
\vskip .3cm
We supersymmetrize a class of moduli dependent potentials
living on branes with the help of additional bulk terms in 5d $N=2$ 
supergravity. 
The space of Poincare invariant vacuum solutions includes 
the Randall-Sundrum solution and the $M$-theoretical solution. 
After adding gauge sectors to the branes we discuss breakdown 
of low energy supersymmetry in this setup and hierarchy of physical scales.
In the limit of large warp factors we find decoupling between effects stemming from different branes in the compactified theory.  


\newpage

\setcounter{page}{1} \pagestyle{plain}
 
The idea of higher-dimensional unification of the fundamental interactions    
attracts considerable interest and receives more and more concrete    
realizations. The general setup for such unification consists of    
hypersurfaces hosting various gauge sectors which are embedded into    
higher dimensional bulk space. Bulk interactions are those of higher   
dimensional gravity coupled to certain scalar and form fields, as well as to   
fermions, which are however inert with respect to gauge groups localized   
on branes.     
{}From the low-energy point of view most of the    
nontrivial features of field theoretical models that are related to    
spatial separation of  gauge sectors should be clearly visible at the level    
of the simple five-dimensional theory. So far, the most extensively studied \cite{ovdw,het,elpp,elp} supersymmetric models of this kind are related to the supergravity model constructed by Horava and Witten as the low energy effective theory of the strongly coupled heterotic $E_8 \times E_8$ superstring in \cite{ew,hw} (see also \cite{fwhw}).  
In particular, a five-dimensional theory is the   simplest nontrivial setup to study spontaneous supersymmetry breakdown and its transmission between the branes \cite{elpp,elp,elpt}. The agents of that transmission are  the bulk fields. 
 
Much attention to five-dimensional gravity has also been drawn by the recent observation that 5d anti-de-Sitter gravity with 3-branes embbedded in the bulk allows for solution with localized gravitational field \cite{rsgrav}. 
However, in this scenario it is necessary to add the  cosmological terms localized on the boundaries with coefficients determined uniquely by the cosmological term in the bulk. The correlation between bulk and boundary potentials is crucial for obtaining a consistent solution to Einstein equations, as well as vanishing of the cosmological constant in the effective four dimensional theory. 
In the original paper \cite{rsscale} no symmetry justifies this apparent fine-tuning. 
 
A considerable effort has been devoted to supersymmetrization of the Randall-Sundrum model, mostly with negative results \cite{kl},\cite{bc}. The exception is the recent reference \cite{bagger}. In that paper the authors start with 5d N=2 pure supergravity with a cosmological constant $\Lambda$ and demonstrate that inclusion of branes in a supersymmetric way leads to the Randall-Sundrum action. Neither scalar fields in the bulk nor gauge and matter on the branes are included in that construction. 
 
The purpose of this paper is to study, in a more general way, a class of  
five dimensional locally supersymmetric theories with 3-branes. We demonstrate that certain types of potentials introduced on  branes, that are not parts of a supersymmetric model on a brane, 
can be supersymmetrized by modifications of the 5d supergravity in the bulk. The requirement of supersymmetry yields relations between bulk and brane cosmological potentials, such that in suitable limits we obtain a supersymmetric version of the Randall-Sundrum scenario or the M-theoretical solution. The general case can still preserve the characteristic features of the Randall-Sundrum model and at the same time 
includes non-trivial potentials for the hypermultiplets of the bulk theory. We find a vacuum solution which preserves one half of the supercharges. This solution can serve as a background for the compactification to the effective 4d theory with N=1 supersymmetry. Next, we include gauge fields on the brane, and discuss the modification of the brane action and supersymmetry
transformation laws, necessary to obtain a supersymmetric theory. \\

{}Finally, we discuss supersymmetry breaking and the role played by the warp factor and conclude the paper with some remarks on the consistency of the compactification to 4d.
   
We start with a five-dimensional N=2 supergravity on the manifold $M_4\times S_1/{\bf Z}_2$ which includes a gravity multiplet $(e_\alpha^m, \psi_\alpha, \ca_\alpha)$ coupled to one hypermultiplet $(\lambda^a, V, \sigma, \xi, \bar{\xi})$ forming a  $SU(2,1)/U(2)$ non-linear sigma model. Two parallel 3-branes are located at $x^5=0$ and $x^5=\pi\rho$. This particular framework is motivated by the Horava-Witten model compactified to 5d \cite{ovdw,elpp}. The sigma-model metric can be read from the K\"ahler potential: 
$ 
K=-ln(S+\bar{S}-2\xi\bar{\xi}), \;
S=V+\xi\bar{\xi}+i\sigma. 
$ 
The conventions and normalizations we use are mainly those of reference \cite{het}. The signature of the metric tensor is $(-++++)$.The SU(2) spinor indices are raised with antisymmetric tensor $\epsilon^{AB}$, and the Sp(1) indices (those of hyperino) with $\Omega^{ab}$.   We choose $\epsilon^{12}=\epsilon_{12}=\Omega^{21}=\Omega_{21}=1$. The rule for dealing with symplectic spinors is $\ov{\psi_1}^A\psi_2^B=\ov{\psi_2}^B\psi_1^A$ 
(note that $\ov{\psi}^A=\ov{\psi_A}$). The  
${\bf Z}_2$ symmetry acts as  reflection $x^5 \rightarrow -x^5$ and is represented in such a way  that bosonic fields $(e_{\mu}^{m},e_{5}^{5}, {\cal A}_{5}, V, \sigma)$ are even, and $(e_{5}^{m},e_{\mu}^{5}, {\cal A}_{\mu}, \xi)$ are odd. The indice $\alpha, \beta ...$  are  five dimensional ($0..3, 5$), while 4d indices are denoted by $\mu, \nu, ...$.The action of the ${\bf{Z_{2}}}$ on fermion fields and on parameter $\epsilon$ of supersymmetry transformations 
is defined as: 
\bea 
\label{z2} 
\gamma_{5}\psi_{\mu}^{A}(x^5)=(\sigma^{3})^{A}\;_{B}\psi_{\mu}^{B}(-x^5)& 
\gamma_{5}\psi_{5}^{A}(x^5)=-(\sigma^{3})^{A}\;_{B}\psi_{5}^{B}(-x^5) 
\nn 
\gamma_{5}\lambda^{a}(x^5)=-(\sigma^{3})^{a}\;_{b}\lambda^{b}(-x^5)& 
\gamma_{5}\epsilon^{A}(x^5)=(\sigma^{3})^{A}\;_{B}\epsilon^{B}(-x^5) 
\eea 
where $\gamma_{5}=(^{-{\bf 1}\;0}_{\;\;\;0\;{\bf 1}}), 
\sigma^{3}=(^{1\;\;\;0}_{0\; -1}), \; A,a=1,2$. Symplectic Majorana spinors  
in 5d satisfy $\bar{\chi}^A = ( C \chi^A)^T$ with $C=-i \gamma^2 \gamma^0$  
in 4d chiral representation. The kinetic part of the action and supersymmetry 
transformation laws up to 3-fermi terms are: 
\beqa 
\label{5daction} 
&S=-\int d^5xe_{5}\frac{1}{2\kaps} \left ( \right .  
R+\frac{3}{2}{\cal F}_{\alpha\beta}{\cal F}^{\alpha\beta}+ 
\frac{1}{\sqrt{2}}\epsilon^{\alpha\beta\gamma\delta\epsilon}{\cal A}_{\alpha} 
{\cal F}_{\beta\gamma}{\cal F}_{\delta\epsilon} 
+\frac{1}{2V^2}(\pa_{\alpha}V\pa^{\alpha}V+D_{\alpha}\sigma D^{\alpha}\sigma) 
& \nonumber \\
&+\frac {2}{V}\pa_{\alpha}\xi\pa^{\alpha}\bar{\xi}
+\frac{i}{2V^2}(\xi\pa_{\alpha}\bar{\xi} D^{\alpha}\sigma-\bar{\xi}\pa_{\alpha}\xi D^{\alpha}\sigma)
-\frac{1}{2V^2}( 
(\xi\pa_{\alpha}\bar{\xi})^2+(\bar{\xi}\pa_{\alpha}\xi)^2 
-|\bar{\xi}\pa_{\alpha}\xi|^2) & \nonumber \\ 
&-(\frac{1}{2}\ov{\psi^1_\mu}\gamma^{\mu\nu\rho}D_\nu\psi^1_\rho + 
 (1\rightarrow 2)) 
-(\frac{1}{2}\ov{\lambda^1}\gamma^\mu D_\mu \lambda^1)+ 
 (1\rightarrow 2)) \left . \right)  & 
\eeqa 
\beqa 
&\delta e_{\alpha}^{m}= 
\frac{1}{2}\ov{\epsilon^1}\gamma^{m}\psi_{\alpha}^1+(1\rightarrow 2)& 
 \nonumber \\ 
&\delta \psi_\alpha^{1}= 
D_{\alpha}\epsilon^{1} 
-\frac{i}{4\sqrt{2}}(\gamma_{\alpha}^{\;\beta\gamma}-4\delta_{\alpha}^{\beta}\gamma^{\gamma}){\cal F}_{\beta\gamma}\epsilon^{1} 
+\frac{i}{4V}D_{\alpha}\sigma\epsilon^{1} 
+\frac{1}{4V}(\xi\pa_{\alpha}\bar{\xi}-\bar{\xi}\pa_{\alpha}\xi)\epsilon^{1} 
-\frac{1}{\sqrt{V}}\pa_\alpha\xi\epsilon^{2} & \nonumber \\ 
&\delta \psi_\alpha^{2}= 
D_{\alpha}\epsilon^{2} 
-\frac{i}{4\sqrt{2}}(\gamma_{\alpha}^{\;\beta\gamma}-4\delta_{\alpha}^{\beta}\gamma^{\gamma}){\cal F}_{\beta\gamma}\epsilon^{2} 
-\frac{i}{4V}D_{\alpha}\sigma\epsilon^{2} 
-\frac{1}{4V}(\xi\pa_{\alpha}\bar{\xi}-\bar{\xi}\pa_{\alpha}\xi)\epsilon^{2} 
+\frac{1}{\sqrt{V}}\pa_\alpha\bar{\xi}\epsilon^{1}  & 
\nonumber \\ 
&\delta{\cal A}_{\alpha}= 
-\frac{i}{2\sqrt{2}}\ov{\psi_\alpha^1}\epsilon^{1}+(1\rightarrow 2)&  
\eeqa 
\beqa 
&\delta V= 
\frac{i}{\sqrt{2}}V(\ov{\epsilon^1}\lambda^1)-(1\rightarrow 2)& 
\nonumber \\ 
&\delta \sigma = 
+\frac{1}{\sqrt{2}}V(\ov{\epsilon^1}\lambda^1)+(1\rightarrow 2) 
+\sqrt{\frac{V}{2}}(\xi\ov{\epsilon^1}\lambda^2-\bar{\xi}\ov{\epsilon^2}\lambda^1)& 
\nonumber \\ 
&\delta \xi= 
-\frac{i\sqrt{V}}{\sqrt{2}}(\ov{\epsilon^2}\lambda^1) 
\; \; \;  
\delta \bar{\xi}= 
-\frac{i\sqrt{V}}{\sqrt{2}}(\ov{\epsilon^1}\lambda^2)& 
\nonumber \\ 
&\delta \lambda^1= 
-\frac{i}{2\sqrt{2}V} (\pa\!\!\!\slash(V+i\sigma) 
-\bar{\xi}\pa\!\!\!\slash\xi+\xi\pa\!\!\!\slash\bar{\xi} ) \epsilon^1 
+\frac{i}{\sqrt{2V}}\pa\!\!\!\slash\xi\epsilon^2& 
\nonumber \\ 
&\delta \lambda^2= 
+\frac{i}{2\sqrt{2}V} ( \pa\!\!\!\slash(V-i\sigma) 
+\bar{\xi}\pa\!\!\!\slash\xi-\xi\pa\!\!\!\slash\bar{\xi} ) \epsilon^2 
+\frac{i}{\sqrt{2V}}\pa\!\!\!\slash\bar{\xi}\epsilon^1 \, .& 
\eeqa  
We assume a scalar potential $\delta(x^5)\frac{e}{\kappa^2}
(-\Lambda+\frac{\sqrt{2}\alpha}{V})$ localized on, say, the first brane (note the delta function), and study the variation of the brane action under supersymmetry transformations. The motivation for the  constant $(\Lambda)$ part of this expression is that it will finally lead us to the Randall-Sundrum expotential solutions. At the same time we allow for cosmological potentials for hypermultiplet scalars; the above form is motivated by the M-theory example and is a natural extension in the presence of hypermultiplets. The generalizations are possible,  but  $\sigma$-dependent terms in the potential break the translational U(1) symmetry $\sigma \rightarrow \sigma+const$ which is useful when we
introduce potential in the bulk, while $\xi$ cannot appear in the boundary potential because of parity assignments.   
 We will be able to supersymmetrize this action by modification of the bulk action only (thus, our construction is alternative to \cite{bagger}).  

{}For simplicity, we initially put $\alpha=0$  and consider a  cosmological term of the form: 
\beq 
\label{brane1} 
\lc_B=-\delta(x^5)\frac{e}{\kaps}\Lambda 
\eeq   
where $e$ is 4d determinant built from the metric induced on the brane. 
We wish to supersymmetrize this term. The supersymmetry variation of $\lc_B$ comes from varying $e$: 
\beq 
\label{bvar} 
\delta\lc=+\frac{1}{2} \delta(x^5)e\Lambda(\ov{\psi_\mu^1}\gamma^\mu\epsilon^1+ 
(1\rightarrow 2)) . 
\eeq 
We observe that, without further modification of the boundary action,  we can cancel this variation by modifying gravitino transformation law: 
\bea 
\label{dpsi} 
\delta \psi_\alpha^1=+\frac{\Lambda}{12}\epsilon(x^5)\gamma_\alpha\epsilon^1 
\nn  
\delta \psi_\alpha^2=-\frac{\Lambda}{12}\epsilon(x^5)\gamma_\alpha\epsilon^2 . 
\eea 
Note that these corrections are compatible with ${\bf Z}_2$ symmetry defined by (\ref{z2}).\\   
If we vary $\psi$ in the gravitino kinetic term, the fifth derivative acting on the step function produces an expression multiplied by a delta function, which  precisely cancels (\ref{bvar}). But now  the bulk theory is not supersymmetric. It is straightforward to show that the variations of the gravitino kinetic term resulting from  (\ref{dpsi}) and proportional to $\Lambda\epsilon(x^5)$ can be cancelled by addding a `gravitino mass term': 
\beq 
\label{gmass} 
\lc_{\psi^2}=+\frac{e_5}{8\kaps}\Lambda\epsilon(x^5) 
( 
\ov{\psi_\alpha^1}\gamma^{\alpha\beta}\psi_\beta^1 
-\ov{\psi_\alpha^2}\gamma^{\alpha\beta}\psi_\beta^2 .
) 
\eeq 
The gravitino variation  $\delta \psi_\alpha^A=D_\alpha \epsilon^A$ in (\ref{gmass}) cancels the above mentioned variation, but now (\ref{dpsi}) applied to the mass term (\ref{gmass}) will produce a variation proportional to $\Lambda^2$, which can be cancelled by varying the determinat in a  new 'cosmological term':\\ 
\beq 
\label{cosmo} 
\lc_{C}=\frac{e_5}{6\kaps}\Lambda^2 .
\eeq 
Moreover, in our framework, $\epsilon(x^5)$ has another discontinuity at $x^5=\pi\rho$ so  an additional term multiplied by $\delta(x^5-\pi\rho)$ appears in the varied bulk Lagrangian. This variation can be cancelled by adding a cosmological term confined to that brane: 
\beq 
\lc_{B'}=\delta(x^5-\pi\rho)\frac{e}{\kaps}\Lambda 
\eeq   
(The minus sign relative to (\ref{brane1}) appears because $\epsilon(x^5)$ has a `step down' at $x^5=\pi\rho$) {}\footnote{The need for the bulk gravitino 
mass term proportional to $\epsilon(x^5)$ has been also pointed out in 
\cite{tgap}.}.\\   
Note that the cosmological term appeared with a plus sign. The relevant part of the bulk action now reads $S=-\frac{1}{2} \int(R-\frac{1}{3}\Lambda^2)$ which allows for anti-de-Sitter solutions. In fact, the coefficient of (\ref{cosmo}) is precisely the one we need to obtain the Randall-Sundrum scenario, as we will show soon.\\  
The above mentioned corrections are still not  
sufficient to supersymmetrize the bulk lagrangian. We also need the hyperino  
mass term: 
\beq 
\lc_{\lambda^2}= 
+\frac{e_5}{8\kaps}\epsilon(x^5)\Lambda \left (\ov{\lambda^1}\lambda^1 
-(1\rightarrow 2)\right ), \eeq 
and the coupling of the graviphoton to gravitino: 
\beq 
\lc_{A}= 
-\frac{ie_5}{4\sqrt{2}\kaps}\epsilon(x^5)\Lambda 
\left ((\ov{\psi^1}_\alpha\gamma^{\alpha\beta\gamma}\psi^1_\gamma){\cal A}_\beta 
 -(1\rightarrow 2)\right ). \eeq 
In addition a graviphoton dependent correction to gravitino transformation  
law appears: \beq 
 \delta \psi^A_\alpha= 
+\frac{i}{2\sqrt{2}}\epsilon(x^5)\Lambda(\sigma^3)^A_{\;B}\epsilon^B\ca_\alpha . 
\eeq 
Further, we need 4-fermi terms in the bulk action  to complete the supersymmetrization, but these are not given in this letter.  
 
Let us now assume $\Lambda=0$ and consider the boundary term: 
\beq 
\label{branealfa} 
\lc=\delta(x^5)\frac{e}{\kaps}\frac{\sqrt{2}\alpha}{V} \,. 
\eeq   
The variation of the determinant can be canceled by modifying $\delta \psi$, similarly to the previous case: 
\bea 
\label{dpsialfa} 
\delta \psi_\alpha^1=-\frac{\sqrt{2}}{12}\frac{\alpha}{V} \epsilon(x^5)\gamma_\alpha\epsilon^1 
\nn  
\delta \psi_\alpha^2=+\frac{\sqrt{2}}{12}\frac{\alpha}{V}\epsilon(x^5)\gamma_\alpha\epsilon^2 \, . 
\eea 
We must also vary the  hyperplet modulus $V$ in (\ref{branealfa}) 
($\delta V=\frac{iV}{\sqrt{2}}(\ov{\epsilon^1}\lambda^1-\ov{\epsilon^2}\lambda^2)$) 
\beq 
\label{bvaralfa} 
\delta\lc=-i\delta(x^5)e\frac{\alpha}{V} ( 
\ov{\epsilon^1}\lambda^1-(1\rightarrow 2)) .
\eeq 
This variation can be cancelled by modifying supersymmetry transformation law of the hyperino $\lambda$: 
\bea 
\label{dlambda} 
\delta \lambda^1=\frac{i}{2V}\alpha\epsilon(x^5)\epsilon^1 
\nn 
\delta \lambda^2=\frac{i}{2V}\alpha\epsilon(x^5)\epsilon^2 \, . 
\eea 
A similar mechanism works: in the variation of the hyperino kinetic term the fifth derivative acts on the step function which leads to a term which precisely cancels (\ref{bvaralfa}). Note that it is only the potential $\alpha/V$ which causes the corrections to the hyperino transformation law. As before, we need to supersymmetrize further. Two-fermi terms and, as a consequence,  a cosmological potential is necessary: 
\beq 
\label{hmass} 
\lc= i \frac{e_5}{2V\kaps}\alpha\epsilon(x^5) 
\left ( 
-\frac{\sqrt{2}}{4}(\ov{\psi^1_\alpha}\gamma^{\alpha\beta}\psi^1_\beta 
-(1\rightarrow 2)) 
+(\ov{\lambda^1}\gamma^\alpha\psi_\alpha^1+(1\rightarrow 2)) 
+\frac{3\sqrt{2}}{4}(\ov{\lambda^1}\lambda^1-(1\rightarrow 2)) 
 \right )
\eeq 
\beq 
\label{cosmoalfa} 
\lc_{C}=-\frac{e_5}{6\kaps}\frac{\alpha^2}{V^2} .
\eeq   
However, this time a minus sign relative to that of (\ref{cosmo}) appears,  
and anti-de-Sitter solution is not allowed. Moreover, contrary to the previous case, 2-fermi and cosmological terms are not enough to render the bulk lagrangian supersymmetric.  
Closer inspection shows, that terms of the form $\alpha(\epsilon\psi)\pa_\alpha\sigma$ do not cancel and the bulk lagrangian must be supplemented with a coupling $\alpha\pa_\beta\sigma\ca^\beta$. In the context of 5d supergravity this means that the translations of the pseudoscalar  
$\sigma$ from  
the hypermultiplet are gauged, with graviphoton being the gauge field. 
To recapitulate, after starting with the  
boundary term (\ref{branealfa}) we are led to 5d gauged supergravity  
similar to that studied in \cite{ovdw}. 
 
One could also imagine other powers of V occuring in (\ref{branealfa}), let  
us say some function $f(V)$. But  
then supersymmetrization is possible only if the bulk sigma model quaternionic metric is found. In some simple cases one can appropriately redefine Re(S)  
and end up in the same sigma model, however in general one has to search for  
new sigma models with quaternionic kinetic metric that allow gauging, 
which is beyond the scope of this paper. 
 
Interestingly enough, we can join both schemes discussed in this paper and  
demand a boundary term: 
\beq 
\label{branejoint} 
\lc_B=\delta(x^5)\frac{e}{\kaps}(-\Lambda+\frac{\sqrt{2}\alpha}{V}) . 
\eeq    
As explained we need a similar term on the second brane: 
\beq 
\label{branejoint2} 
\lc_{B'}=-\delta(x^5-\pi\rho)
\frac{e}{\kaps}(-\Lambda+\frac{\sqrt{2}\alpha}{V}) . 
\eeq    
Repeating the same line of arguments, we arrive at the conlusion, that we need the gauged supergravity in the bulk of the kind considered in \cite{ovdw}, but with the potential: 
\beq 
\label{cosmojoint} 
\lc_{C}=\frac{e_5}{6\kaps}(-\Lambda+\frac{\sqrt{2}\alpha}{V})^2 
-\frac{e_5}{2\kaps}\frac{\alpha^2}{V^2} \, . 
\eeq   
Additional terms are needed to arrive at completely supersymmetric  
bulk action and they all fit into the general form of gauged supergravity 
with local translations of $\sigma$.  
 
Since we want to compactify this theory down to 4d and demand that the  
effective theory has N=1 supersymmetry, we must search for the background which preserves exactly four supercharges. The supersymmetry transformation laws of fermions, including modifications found in the previous paragraphs are:   
\bea 
\delta \psi_\alpha^A= 
 D_\alpha \epsilon^A 
-\epsilon(x^5)\frac{1}{12}(-\Lambda+\frac{\sqrt{2}\alpha}{V})\gamma_\alpha 
(\sigma^3)^A_{\;B}\epsilon^B 
\nn 
\delta \lambda^a= - 
 \frac{i}{2\sqrt{2}V}\pa_5V\gamma^5(\sigma^3)^a_{\;B}\epsilon^B 
 + \alpha\epsilon(x^5)\frac{i}{2V}\epsilon^a \, . 
\eea 
In the above formulas we neglected terms with 4d derivatives $\pa_\mu$ so as to preserve 4d Poincare invariance. We also put $\sigma=\ca_5=0$ since these fields do not occur in the potential, so this choice is consistent with  equations  of motion. Finally, we neglected $\pa_5 \xi$ term since, as we show later in this letter, expectation value of this term generically leads to supersymmetry breaking.  
 
The ansatz for static solutions is: 
$ 
ds^2=a(x^5)dx^\mu dx^\nu \eta_{\mu\nu}+b(x^5)(dx^5)^2, \; V=V(x^5) 
$. 
The relevant supersymmetry transormation laws evaluated for this ansatz are ( ' denotes $\pa_5$ and the world indices are with respect to the Minkowski metric $\eta$): 
\beqa 
&\delta \psi_\mu^A= 
\frac{a'}{4\sqrt{ab}}\gamma_\mu \gamma_5 \epsilon^A 
-\epsilon(x^5)\frac{\sqrt{a}}{12}(-\Lambda+\frac{\sqrt{2}\alpha}{V})\gamma_\mu 
(\sigma^3)^A_{\;B}\epsilon^B& 
\nonumber \\ 
&\delta \psi_5^A=\pa_5\epsilon^A- 
\epsilon(x^5)\frac{\sqrt{b}}{12}(-\Lambda+\frac{\sqrt{2}\alpha}{V})\gamma_5 
(\sigma^3)^A_{\;B}\epsilon^B&  
\nonumber \\ 
&\delta \lambda^a= - 
 \frac{i}{2\sqrt{2b}V}V'(\sigma^3)^a_{\;B}\gamma_5 \epsilon^B 
 + \alpha\epsilon(x_5)\frac{i}{2V}\epsilon^a .&  
\eeqa 
We obtain conditions for unbroken supersymmetry by demanding that the above  
variations of fermionic fields are vanishing for vacuum configurations 
\beqa 
\label{rruchu} 
&\frac{a'}{a}=\frac{1}{3}(-\Lambda+\frac{\sqrt{2}\alpha}{V})\epsilon(x^5) 
\sqrt{b} 
& \nonumber \\ 
&V'=\sqrt{2}\alpha\epsilon(x^5)\sqrt{b}& 
\nonumber \\ 
&\pa_5\epsilon^A=\frac{\sqrt{b}}{12}(-\Lambda+\frac{\sqrt{2}\alpha}{V})\epsilon(x^5)\epsilon^A .& 
\eeqa 
In addition we need chirality  conditions for the supersymmetry generating spinor , which  
reduce  N=2 supersymmetry down to N=1: 
\bea 
\label{chir} 
\gamma_5\epsilon^1=\epsilon^1 
& 
\gamma_5\epsilon^2=-\epsilon^2 
\eea 
It turns out, that if the parameters $a,\, b,\, V$ of our ansatz satisfy conditions (\ref{rruchu}), they automatically satisfy the equations of motion (with delta sources), and give vanishing vacuum energy. We can easily solve the conditions (\ref{rruchu}). In the coordinate frame where $b=R_0^2$ the 
 vacuum
solution is: 
\beqa 
\label{vacuum}
&V=V_0+\alpha\sqrt{2}\, R_0 \, (|x^5|-\frac{\pi\rho}{2})& \nonumber \\ 
&g_{\mu\nu}= \left ( 1+\alpha\sqrt{2}\frac{R_0}{V_0}(|x^5|-\frac{\pi\rho}{2}) \right )^{1/3} 
e^{\frac{-R_0 \Lambda}{3}|x^5|}\eta_{\mu\nu}& \nonumber \\ 
& g_{55} = R^{2}_{0}& 
\eeqa 
where a constant coefficient in the solution for $g_{\mu\nu}$ has been absorbed into a redefinition of the relation between 5d and 4d Planck scales.
With the standard procedure we identify the four-dimensional Planck scale as
\beq 
M^{2}_4 = 2 M^{3}_5 R_0 \int_{0}^{\pi \rho} dx^5 a(x^5) =  2 M^{3}_5 R_0 
\int_0^{\pi\rho}dx^5 
 \left (1+\alpha\sqrt{2}R_0(|x^5|-\frac{\pi\rho}{2}) \right )^{1/3} 
exp(-\frac{R_0}{3}\Lambda x^5 ) 
\eeq 
In particular, for $\alpha=0$ one obtains $M^{2}_4 =  
\frac{6 M^{3}_5}{\Lambda} (1 - e^{-\frac{\Lambda R_0 \pi \rho} 
{3}})$.    
 
{}For $\Lambda \rightarrow 0$ we are back in the domain wall solution studied in \cite{ovdw}, while in the case $\alpha \rightarrow 0$ we get the Randall-Sundrum expotential solution  (the connection with the normalization of reference \cite{rsgrav} is $\Lambda=6k$). If we assume $\alpha\rho$ to be small (which is the case in the M-theoretical scenario) we are very close to the Randall-Sundrum solution and, in particular, gravity is still localized on the positive tension brane at $x^5=0$.    
 
{}For phenomenological applications we need gauge and charged matter fields transforming in representations of the Standard Model. The problem of coupling confined to a boundary gauge and matter fields to 5d supergravity can be studied 
along the lines of the original Horava-Witten procedure \cite{hw} and details 
of this can be found in \cite{falk}. 
We summarize the results. Let us add a gauge multiplet $(A_\mu^a, \chi^a)$ , say, on the first brane ('a' is the group index) and set the kinetic function to $-\frac{V}{4}F_{\mu\nu}F^{\mu\nu}$. It turns out that supersymmetric coupling is possible and no changes in the bulk lagrangian are required. All we need is to add the boundary Lagrangian ($g^2$ is the reference gauge coupling): 
\beqa 
\label{YM} 
&{\cal L}_{YM}=\frac{e_4 \delta(x^5)}{g^2} 
\left ( \right .  
 -\frac {V} {4} 
F_{\mu\nu}^{a} F^{a\mu\nu}   
-\frac{1}{4} 
\sigma F_{\mu\nu}^{a} \tilde F^{a\mu\nu} 
-\frac {V} {2} 
\overline{\chi^{a}}D\!\!\!\!\slash\chi^{a} 
+\frac {V} {4}  
(\overline{\psi}_{\mu}\gamma^{\nu\rho}\gamma^{\mu}\chi^{a})F^{a}_{\nu\rho} 
& \nonumber \\
&+\frac {3i} {4\sqrt{2}}\frac {V}{e_{5}^{5}} 
(\overline{\chi}^{a}\gamma^{5}\gamma^{\mu}\chi^{a}){\cal F}_{\mu 5}
-\frac {1} {4} 
(\overline{\lambda}\gamma^{\nu\rho}\chi^{a})F_{\nu\rho}^{a} 
-\frac {i} {8} 
(\overline{\chi}^{a}\gamma^{5}\gamma^{\mu}\chi^{a})\partial_{\mu}\sigma 
& \nonumber \\
&-\frac{\sqrt{V}}{4e_{5}^{5}} 
( 
(\overline{\chi^{a}}_{L}\chi^{a}_{R})\partial_{5}\overline{\xi} 
+ (\overline{\chi^{a}}_{R}\chi^{a}_{L})\partial_{5}\xi 
) 
+(4fermi) \left . \right ).& 
\eeqa  
In the above the bulk fermions appear in their even (and Majorana in the 4d sense) combinations defined as:  
\bea 
\label{spinors}  
 \psi_{\mu} = \left ( 
\begin{array}{cccc} 
{i\psi_{L\mu}^{2}} \\ 
 {i\psi_{R\mu}^{1}} 
\end{array} 
\right ) 
& 
 \psi_{5} = \left ( 
\begin{array}{cccc} 
{-i\psi_{L5}^{1}} \\ 
 {i\psi_{R5}^{2}} 
\end{array} 
\right ) 
& 
 \lambda =\sqrt{2}V \left ( 
\begin{array}{cccc} 
{-\lambda_{L}^{1}} \\ 
 {\lambda_{R}^{2}} 
\end{array} 
\right ) . 
\eea   
We also need to modify the supersymmetry transformation laws of the even bulk fermions: 
\beqa 
&\delta\psi_{\mu}=\delta(x^{5})\frac{\kaps}{g^2}\frac{V}{8} 
(g^{\mu\rho}-\frac{1}{2}\gamma^{\mu\rho})\gamma^{5}\epsilon\; (\overline{\chi^{a}} 
 \gamma^{5}\gamma_{\rho}{\chi^{a}})& \nonumber \\ 
&\delta\lambda= 
\delta(x^{5})\frac{\kappa^{2}}{g^2}\frac{V^{2}}{4} 
\left ( 
\epsilon(\overline{\chi}^{a}\chi^{a})-\gamma^{5}\epsilon(\overline{\chi}^{a}\gamma^{5}\chi^{a}) \right ) .& 
\eeqa 
In the above, the supersymmetry parametr $\epsilon$ is defined as an even combination of 5d supersymmetry parameters:
\beq
 \epsilon = \left (
\begin{array}{cccc}
{i\epsilon_{L}^{2}} \\
 {i\epsilon_{R}^{1}}
\end{array}
\right ).
\eeq 
Note that no gaugino dependent correction appears in the transformation law of $\psi_5$.\\  
In analogy to heterotic models, we have the possibility to break supersymmetry by gaugino condesation on the hidden and/or visible brane. The supersymmetry breaking is transmitted between branes by the expectation value of the hypermultiplet field $\xi$. This mechanism arises because $\xi$, although odd, couples to gauginos on the boundaries through its fifth derivative. The equation of motion for $\xi$ in the presence of the condensates is: 
\bea 
\frac{1}{\kaps}\pa_5(\frac{e_5 g^{55}}{V}\pa_5\xi)= 
\pa_5 \left ( 
-\frac{e_4\sqrt{V}}{2g^2 e_5^5} ( 
\delta(x^5)(\ov{\chi}_L\chi_R)_1+ 
\delta(x^5-\pi\rho)(\ov{\chi}_L\chi_R)_2 
                                ) \right ).
\eea 
We are interested in the solution for $\pa_5 \xi$ because this expression (and not $\xi$ alone) appears in the relevant formulae.  
For $\alpha=0$ the solution is : 
\beq 
\pa_5 \xi= 
-\frac{\kaps}{2g^2}V_0^{3/2} \left (\delta(x^5)\chi_1^2+\delta(x^5-\pi\rho)\chi_2^2 \right ) 
+C exp(\frac{2R_0}{3}\Lambda|y|) .
\eeq 
The non-trivial background effects are due to the expotential factors. The 
constant C can be determined from the boundary conditions (in other words, from matching delta singularities in the equation of motion): 
\beq 
C=\frac{R_0 \Lambda}{e^{\frac{2R_0}{3}\Lambda\pi\rho}-1}  
\frac{\kaps}{ 6 g^2}V_0^{3/2} 
\left (  
\chi_1^2+\chi_2^2 
\right ).
\eeq 
For $\Lambda=0$ it is customary \cite{ovdw} to go to a different coordinate frame  
where $g_{55}=R_0^2 H^4$, $g_{\mu\nu}=\frac{1}{R_0}H\bar{g}_{\mu\nu}, V=V_0 H^3$ and $H=1+\frac{\alpha\sqrt{2}\pi\rho R_0 }{3 V_0 }
(|x^5|-\frac{\pi\rho}{2})$.  
In this frame  we obtain the solution: 
\beq 
\label{dksi} 
\pa^5 \xi H^{-3}= 
\frac{\kaps}{2g^2}V_0^{3/2}H^{3/2}\left (
-\delta(x^5)\chi_1^2 
-\delta(x^5-\pi\rho)\chi_2^2 \right ) 
+C 
\eeq 
\beq 
C= 
\frac{\kaps}{3g^2}\alpha\sqrt{2}\pi\rho V_0^{3/2} 
\frac{ 
-\chi_1^2 H^{9/2}(0) 
-\chi_2^2 H^{9/2}(\pi\rho) 
     } 
{H^4(0)-H^4(\pi\rho)} .
\eeq 
It is worth noting, that in the 5d theory gaugino condensates break supersymmetry (but, if we assume superpotentials on the branes we can cancel their contribution).  In the presence of the condensates we have no way to satisfy simultanously $\delta\psi_\mu^A=0$ and neither of the remaining conditions for unbroken supersymmetry.  Indeed, $\pa_5 \xi$ and condensates do not alter the transformation law of $\psi_\mu$, so in particular, the conditions resulting from $\delta\psi_\mu^A=0$ include chirality conditions (\ref{chir}). But then, the condensates in $\delta \lambda^a$ and $\delta \psi_5^A$ multiply the supersymmetry parameter $\epsilon$, which is of the chirality opposite to  other $\epsilon$'s occuring in these transformation laws. Thus, conditions $\delta\psi_5^A=0$ and   $\delta \lambda^a=0$ cannot be satisfied.

When we compactify our model to 4d on the background (\ref{vacuum}), the independent of $x^5$ integration constants $R_0, V_0$ together with zero modes of $\sigma$ and $\ca_5$,  become the ($x^\mu$ dependent) moduli of the effective 4d theory. The fluctuations around Minkowski metric in the solution (\ref{vacuum}) are described by $\bar{g}_{\mu\nu}$ 
To go to the 4d Einstein frame one needs to perform explicit integration over $x^5$ and a suitable moduli dependent Weyl rotation. We rescale the metric 
$\bar{g}_{\mu\nu} \rightarrow a_0\bar{g}_{\mu\nu}$ with $a_0$ chosen (up to a numerical, independent of moduli, factor which can be absorbed into the definition of the 4d gravitational constant) as 
$a_0^{-1}=\int_0^{2\pi\rho}dx^5 a(x^5)$.
The Killing spinors generating an unbroken $N=1$ supersymmetry are: 
\bea 
\label{killing} 
\epsilon^{1}_{R} = e^{-\frac{\Lambda R_0|x^5|}{12}}  
 \left (1 + \alpha \sqrt{2}\frac{R_0}{V_0} (|x^5| - \frac{\pi \rho}{2}) \right )^{1/12} 
 a_0^{-1/4} \eta_R 
\nn 
\epsilon^{2}_{L} = e^{-\frac{\Lambda R_0 |x^5|}{12}}  
 \left (1 + \alpha \sqrt{2}\frac{R_0}{V_0} (|x^5| - \frac{\pi \rho}{2}) \right )^{1/12} 
 a_0^{-1/4} \eta_L .
\eea 
 
Since $\epsilon^{2}_L = - i \sigma^2 \epsilon^{1 \, *}_R$ (5d Majorana condition) 
spinor $\eta$ is Majorana in the 4d sense. 
The factor $a_0$ in (\ref{killing}) yields canonical form of the reduced 
4d supersymmetry transformation law of the gravity multiplet, $\eta$ depends only on $x^\mu$ and  has an interpretation of  a parameter of supersymmetry transformations 
in the 4d theory. 

When $\Lambda \neq 0$ it is pretty difficult to compactify the
 supersymmetric model which we have defined to four dimensions and even more 
to bring it into the standard 4d supergravity form. 
Hence we postpone the discussion of this case for a while and discuss first 
in 
detail the simpler case with $\Lambda=0$ and $\alpha \neq 0$. 
We can now proceed with the derivation of the 4d effective theory. In that  case the K\"ahler potental is \cite{ovdw}:\\ 
\beq 
\label{kahler} 
K=-ln(S+\bar{S}) 
-3\, ln(T+\bar{T}) .
\eeq 
The moduli $S,T$  and their superpartners are defined as: 
\bea 
\label{moduli} 
S=V_0+i\sigma_0 
& 
\Lambda^S=<(\frac{H}{R_0})^{1/4}(\lambda-\alpha\sqrt{2}(|x^5|-\pi\rho)\psi_5)> 
\nn 
T=R_0+i\sqrt{2}\ca_5 
& 
\Lambda^T=<(\frac{H}{R_0})^{1/4}\psi_{5}> 
\eea 
where $< (...) > = \frac{1}{2 \pi \rho} \int \, dx^5 \, (...)$. 
The gauge sectors originating from two different branes are described by the gauge functions, which includes corrections linear in $\alpha$:  
\bea 
\label{f} 
f_1=S-\frac{\sqrt{2}}{2}\alpha\pi\rho \, T 
\nn  
f_2=S+\frac{\sqrt{2}}{2}\alpha\pi\rho \, T .  
\eea
 The physical gauginos can be expressed as
 $(\chi_1)_{p}=(\chi_1) (\frac{H(0)}{R_0})^{3/4}$,   
$(\chi_2)_{p}=(\chi_2)(\frac{H(\pi\rho)}{R_0})^{3/4}$.

The above corrections were extracted from the kinetic terms of the 5d lagrangian compactified to 4d. Although the functions K and f are sufficient to reconstruct the rest of the supergravity lagrangian, an interesting consistency check would be to obtain explicitly the complete 4d lagrangian by integrating out the fifth dimension. This is fairly difficult as, e.g., the 4-fermi terms have higher order in $\alpha$ contributions. Another approach is to reduce 5d supersymmetry transformation laws to 4d, and check if they are consistent with the results (\ref{kahler},\ref{f}). 
 This has the advantage that corrections can be seen  at lower order in the expansion in $\alpha$ and $\kaps$. As an example we present how to determine the gauge kinetic functions from the transformation laws of moduli superpartners. 
We use the definition of $S$ and $T$ superpartners (\ref{moduli}) and substitute $\pa_5 \xi$ with the solution of its equation of motion in the relevant part of 5d supersymmetry transformation law of $\lambda$ and $\psi_5$. After integrating  
over fifth dimension the result up to $\alpha^2$ corrections is: 
\bea 
\label{fermisusy}
\delta \Lambda^S_L= 
\frac{\kaps_4}{2g^2} V_0^2 \left (\chi_1^2+\chi_2^2 \right ) \eta_L 
\nn 
\delta \Lambda^T_L= 
-\frac{\kaps_4}{12g^2}R_0^2\alpha\sqrt{2}\pi\rho \left (\chi_1^2-\chi_2^2 
\right ) \eta_L \, . 
\eea 
Noting that in 4d supergravity, scalar gaugino condensates in the transformation law of the fermions $\Lambda^S,\Lambda^T$ are multiplied by $\frac{1}{8}f,_S(K^{-1})^S_S$ and $\frac{1}{8}f,_T(K^{-1})^T_T$, respectively, the result indeed agrees with (\ref{f}). A noteworthy detail in this derivation is that in 5d $\pa_5\xi$  appears as a full square: $\partial_{5}\hat{\xi}=\partial_{5}\xi 
+\frac{\kappa^{2}}{g^2}\delta(x^{5})\frac{V^{3/2}}{2} 
((\ov{\chi_R}\chi_L)_1+(\ov{\chi_R}\chi_L)_2)$ in $\delta \lambda$ but not in  
$\delta \psi_5$. Thus, when we calculate $\delta \Lambda^T$ the linear part of the solution for $\pa_5\xi$ cancels to zeroth order in $\alpha$  with delta functions occuring in this solution, leading to the correct form of $f_{,T}$. Note also, that the admixture of $\psi_5$ in the definition of $\Lambda^S$ is crucial to obtain the correct form of  $f_{,S}$.  These conclusions confirm fully the results of \cite{elpp,elp,elpt}.\\
From the transformation laws (\ref{fermisusy}) it can be read off that presence of gaugino condensates breaks supersymmetry also in the 4d effective theory. Although one can adjust $\chi_1^2=-\chi_2^2$ so that the condesates cancel in the regular part of the solution (\ref{dksi}) for $\pa_5 \xi$ and in consequence in $\delta \Lambda^S$, but then the non-zero condensate contribution appears in $\delta \Lambda^T$ due to the above mentioned lacking of the `full square' structure of $\delta \psi_5$. However, if we allow for boundary scalar fields, by appropriate adjusting of their superpotentials we have the possibility to cancell the contribution of the condensates.   

To be precise, we note that the above cosiderations for the case $\Lambda=0$ are valid only to linear order in $\alpha$. In $\alpha^n$ order, with $n > 1$, further corrections appear, but they are difficult to calculate.  
One needs to solve the equations of motion for KK modes of the bulk fields (to linear order in $\alpha$ it suffices to know the expectation value of the bulk fields on the branes). 


While neglecting the higher order corrections in $\alpha$ can be justified by the expected smallness the expansion parameter $\alpha\pi\rho$, this is not  
the case for $\Lambda\pi\rho$, since $\Lambda$ is expected to be set by the string scale.  
Thus, finding the effective theory with a non-zero $\Lambda$, even for the case $\alpha=0$, requires more elaborate tools.          
However, although we do not know the complete effective Lagrangian, 
we can still try to extract some information about the low energy 4d theory by computing 
physically important 4d operators, and using experience gained in the study of the simpler model. Let us put $\alpha =0$ in what follows. 
We know already that the parameter controlling the breakdown of low energy 
supersymmetry is  the regular part of $\partial_5 \xi$
given in formula (\ref{dksi}). This equals in the present case 
$
\partial_5 \xi (x^5) = \frac{1}{e^{\frac{2}{3}R_0\Lambda\pi \rho} -1} 
\frac{ \kappa^{2} R_0 \Lambda }{6 g^2} V_{0}^{3/2} (\chi^{2}_1 
+ \chi^{2}_2)e^{\frac{2}{3}R_0\Lambda |x^5|}.
$\\
We note that in the presence of gaugino codensates the vacuum expectation 
value of $\pa_5 \xi$  is nonzero, and is modulated 
by an expotential, $x^5$-dependent factor. 
The above expression  can be inserted  into 
the bulk kinetic term of $\pa_5 \xi$ (the singular part of the solution drops out due to the `full square' structure), to read off physical gaugino masses on each wall. These are 
the important parameters as they can tell us directly the physical 
magnitudes of induced global 
supersymmetry breaking terms in the boundary gauge sectors. 
In the limiting case $\Lambda \pi \rho \ll 1$, to the lowest order (and keeping
$\alpha = 0$), 
we recover this way soft gaugino masses which we have obtained in the 
compactification of the pure M-theoretical model. 
To perform the task in a general case, we need to expand metric around the 
vacuum solution. 
At the same time, to obtain kanonical normalization of kinetic terms of fields living on the 
branes we need to rescale them by expotential factors. This also gives  the standard form of the  
 supersymmetry transformation law of the gauge field $A_\mu$ reduced in our 
background,
   $\delta A_\mu=\ov{\eta}\gamma_\mu\chi$. The needed rescaling is 
$\chi_i = a_i^{-3/4} (\chi_{i})_{p}$, i.e.   
$  
(\chi_1)_{5d}=a_0^{-3/4}(\chi_1)_{p}$ and $  
(\chi_2)_{5d} = a_0^{-3/4}e^{\frac{1}{4}R_0\Lambda\pi\rho}(\chi_2)_{p}$. 
If we interpret quartic gaugino terms as leading to gaugino masses after condensation the result for the masses is (in the limit $\Lambda \pi \rho 
\gg 1$) \footnote{It is useful to express the 5d gravitational coupling $\kappa^2$ through the 
4d coupling $\kappa^{2}_{4}$. In the two cases of interest the relation 
is a) $\kappa^{2}_{4} = \kappa^2 / ( 2 \pi \rho )$ when $\Lambda \pi \rho 
\ll 1$ and b) $\kappa^{2}_{4} = \kappa^2  \Lambda / 6$ when 
$\Lambda \pi \rho \gg 1$.}.
\beqa
&M_1 = \frac{V_{0} \kappa_{4}^2 }{2 g^4} 
e^{-\frac{2}{3}R_0\Lambda\pi \rho}
\left ( 
<\chi_{1 \, p}^{2}> 
+ <\chi_{2 \,p}^2> e^{\frac{1}{2}R_0\Lambda\pi \rho}\right )& \nonumber \\
&M_2 = \frac{V_{0} \kappa_{4}^2 }{2 g^4} 
e^{-\frac{1}{6} R_0\Lambda\pi \rho}
\left ( 
<\chi_{1 \, p}^{2}> 
+ <\chi_{2 \,p}^2> e^{\frac{1}{2}R_0\Lambda\pi \rho}\right )
.&
\eeqa
For definitness of the discussion, let us consider one after another two
possible visible-hidden sector configurations. 
First we put the visible sector on the negative tension brane at $x^5 = \pi \rho$ and switch on the hidden condensate on the positive tension brane. 
First of all, we see that $M_2 \approx e^{-\frac{1}{6}R_0\Lambda\pi \rho} < \chi_{1 \, p}^2 > $ becomes exponentially suppressed as a function of the distance between walls, and it scales exactly as expected on the basis of the 
argument given in \cite{rsscale}. 
If we now revert the roles of the branes, and assume that the positive tension brane contains observable fields, the situation is similar, i.e. 
$M_1 \approx e^{-\frac{1}{6}R_0\Lambda\pi \rho} < \chi_{2 \, p}^2 > $ vanishes with growing distance between the walls. The thing to be noted in this case, 
which can be called 
standard hidden sector scenario, is the presence of the additional exponential 
factor in front of the usual dynamically generated mass scale $ 
a^{1/2} (\pi \rho) \Lambda^{3}_{hid} / M^{2}_{Pl}$, which is the (not necessarily welcome) 
source of additional hierarchy. In conclusion we stress that in the limit of large warp factors the transmission of supersymmetry breaking from the hidden brane is expotentially suppresed. Thus, in this scenario walls indeed decouple with the growing distance between them.      \\
A different role of the warp factor can be seen when one considers a
condensate forming on the same wall where the observable sector lives. 
If this happens on the negative tension wall the induced gaugino mass 
seems to be exponentially enhanced, like $e^{\frac{1}{3}R_0\Lambda\pi\rho}$. If on the other hand 
the visible brane is the positive tension one, the supersymmetry breaking mass  is suppressed by the factor $e^{-\frac{2}{3}R_0\Lambda\pi\rho}$.
 One obvious comment
on this is that the walls are not equivalent, in the sense of being interchangable, 
as was already the case in the nonsupersymmetric Randall-Sundrum scenario. 
Second, let us note that expecting large, say well above $1 \; TeV^3$,
condensates on the negative tension brane where  all 
physical scales are scaled down 
to say $1 \; TeV$ may be inconsistent in the present framework. 

At this point we should consider matter fields on the boundaries, interacting 
by means of a trilinear superpotential $W$. New terms in the Lagrangian which 
should be taken into account are
\beqa
&S_{scalar} = \int d^5 x \frac{ e_4}{g^2} \delta (x^5 ) 
\left ( - D_\mu \Phi D^\mu \bar{ \Phi}
- \frac{2}{V} \frac{\partial W}{\partial \Phi } \frac{\partial \bar{W}}{\partial \bar{\Phi}} - \frac{4 \kappa^2 }{V} W \bar{W} + \frac{2}{V e^{5}_{5}} W \partial_5 \xi + \; h.c.\; \right )& 
\eeqa
(and similarly  for the second wall).  
The results of the coupling between $W$ and $\partial_5 \xi$ are twofold. 
First, in the previous formulae for the vacuum solution for $\partial_5 \xi $ 
one should substitute 
\beq
\chi_i \bar{\chi}_i \rightarrow  \chi_i \bar{\chi}_i - \frac{4 W_i}{V^{3/2}}.
\eeq
This means that expectation value of $W$ contributes to the supersymmetry breaking, and can in principle cancel the contribution of condensates. 
Let us note, that the canonical normalization of kinetic terms of scalars 
$\Phi_i$ living on the $i$-th wall leads to rescaling $\Phi_1 \rightarrow 
\Phi_{1 \, p} a_0^{-1/2}$, $\Phi_2 \rightarrow \Phi_{2 \, p} a_0^{-1/2}e^{\frac{1}{6}R_0\Lambda\pi\rho}$. This implies that superpotential from the $i$-th wall scales like 
a condensate from the same wall, hence the earlier discussion of decoupling 
applies here without modifications.\\
The second result of new couplings is the appearance of softly breaking 
global supersymmetry trilinear scalar terms when condensates 
are switched on. These terms, say on the second brane,  
are proportional to  
\beq 
a^{1/2}_2 \frac{1}{V e^{5}_{5}} W_{2 \, p}  <\chi^1_{p} - \frac{4 W_{1 \, p}}{V^{3/2}} > 
\eeq
and one can easily work out their scaling properties. These terms are the 
physical soft terms assuming that the effective 4d vacuum energy, after 
switching on vevs for boundary scalars, vanishes.

After presenting this preliminary 
and somewhat speculative interpretation of the supersymmetry breaking 
pattern, we want to stress that
definite conclusions can be made only when one constructs the complete 
effective 4d theory.

To put this discussion into a wider framework, let us remind ourselves 
that what we have done here so far is the traditional 
compactification of the fifth dimension, where one tries to combine both,
in principle different, gauge sectors at the  
ends of the five-dimensional world into a single effective theory. 
However, a different approach is possible, see \cite{verlinde}. 
One can imagine that the model living on the negative tension brane located at 
$x^5 = \pi \rho$ 
is a holographic image of the same model living on the Planck brane located 
at $x^5 =0$. Flow of the second brane along the $x^5$ axis accompanied 
by rescaling  of  all the mass scales on that brane by a factor $a^{1/2} (x^5)$
might be considered to be equivalent to renormalization group flow along 
momentum scale towards the IR limit. In this context we want to notice, that 
in the $N=1$ theory which we study here the mass scales scale exactly in 
the way required by holographic principle, but the gauge coupling 
does not scale with the changing warp factor. This is easy to see, since 
the warp factor cancels out from the expression $ e_4 g^{\mu \nu} g^{\beta \alpha} F_{\mu \beta} F_{\nu \alpha}$. The intriguing observation is that if one 
would try to improve for that, and scale also the gauge coupling according to 
one-loop scaling anomaly, see \cite{ddg}, $\frac{1}{g^2 (x^5)} = \frac{1}{g^2 (M_5) } + b_0 \log \left (
\frac{M_5}{m (x^5)} \right )^2 = \frac{1}{g^2 (M_5) } - b_0 \log a(x^5) $,
then assuming $\frac{1}{g^2 (x^5)}=\frac{1}{g^{2}_{GUT} (x^5)}$ this flow 
would compensate the relative enhancement factor between gaugino condensate 
from the second and first wall, $\Lambda_{cond} (\pi \rho) = M_5 e^{-\frac{1}
{2 b_0} (\frac{1}{g^2} - b_0 \log (a(\pi \rho )))} \approx a^{1/2} (\pi \rho )
\Lambda_{cond} (0)$.  

At the end we would like to comment on two aspects of the models we discuss in this paper. Firstly,
it is interesting to note that the additional terms which we have put on the boundary, $\delta L = e f(V)$, are not parts of a globally supersymmetric sigma model  
living on a brane. When one integrates over the fifth dimension these terms cancel against the bulk potential and drop out completely from the effective  
four dimensional model. Hence, the presence of the additional dimension offers 
the possibility of supersymmetrizing certain boundary terms  
along the direction transverse to the branes, with partner terms living  
in the bulk. Secondly, the vanishing vacuum energy in the pure bulk moduli  
sector which we observe does not solve automatically the cosmological 
constant  
problem. When we allow matter chiral superfields on the branes to follow their local dynamics given by nontrivial superpotential and gauge interactions, the 
new vacuum they approch is not guaranteed to give automatically a 
vanishing contribution to the 4d vacuum energy, and in general next instance 
of tuning is necessary.

To summarize, we have presented a class of five dimensional supergravities 
with gauge sectors living on 4d boundaries, which admit exponential 
warp factors analogous to that of the Randall-Sundrum model. The required fine-tuning between bulk and boundary cosmological potetnials was explained by supersymmetry.  
These models can be considered to be deformations of the M-theoretical 
model constructed in \cite{ovdw}. We have discussed  hidden sector 
supersymmetry breaking and its transmission between branes in the present 
models. The setup and results are likely to be relevant for the discussion 
of the holographic projection of $N=1$ supersymmetric gauge models.

\vspace{1.5cm}
\noindent This work has been supported by TMR programs
ERBFMRX--CT96--0045 and CT96--0090.
Z.L. and S.P. are supported
by the Polish Committee for Scientific Research grant 2 P03B 05216(99-2000).


\begin{thebibliography}{99} 
\bibitem{ovdw} A. Lukas, 
B. A. Ovrut, K. S. Stelle and D. Waldram, {\it Phys. Rev.} {\bf D59} (1999) 
086001. 
\bibitem{het} A. Lukas, B. A. Ovrut, K. S. Stelle and 
D. Waldram, {\it Nucl. Phys.} {\bf B552} (1999) 246. 
\bibitem{elpp} J. Ellis, Z. Lalak, S. Pokorski, W. Pokorski, {\it Nucl.Phys.}
 {\bf B540} (1999) 149. 
\bibitem{elp} J. Ellis, Z. Lalak, W. Pokorski, {\it Nucl.Phys.} {\bf B559} 
(1999) 71. 
\bibitem{ew} E. Witten, {\it Nucl.Phys.} {\bf B471} (1996) 135.
\bibitem{hw} P. Horava and E. Witten, {\it Nucl.Phys.} {\bf B460} (1996) 
506 and {\it Nucl.Phys.} {\bf B475} (1996) 94. 
\bibitem{fwhw} T. Banks, M. Dine, {\it Nucl. Phys.} {\bf B479} 
(1996) 173;
H.P. Nilles, S. Stieberger, {\it Nucl. Phys.} {\bf B499} (1997) 3;
I. Antoniadis, M. Quiros, {\it Nucl. Phys.} {\bf B505} (1997) 109;
T. Li, J. L. Lopez, D.V. Nanopoulos, {\it Phys. Rev.} {\bf D56} 
(1997) 2602;
E. Dudas, Ch. Grojean, {\it Nucl. Phys.} {\bf B507}
(1997) 553;
H.P. Nilles, M. Olechowski, M. Yamaguchi,
{\it Phys. Lett.} {\bf B415} (1997) 24;
Z. Lalak, S. Thomas, {\it Nucl. Phys.} {\bf B515} (1998) 55;
A. Lukas, B. Ovrut, D. Waldram,
{\it Nucl. Phys.}  {\bf B532} (1998) 43;
K. Choi, H. B. Kim, C. Munoz, {\it Phys. Rev.} {\bf D57} (1998) 7521. 
\bibitem{elpt} J. Ellis, Z. Lalak, S. Pokorski, S. Thomas, {\it Nucl.Phys.}
{\bf  B563} (1999) 107. 
\bibitem{rsgrav} L. Randall and R. Sundrum, {\it Phys. Rev. Lett.} {\bf 83} (1999) 4690.
\bibitem{rsscale} L. Randall and R. Sundrum, {\it Phys. Rev. Lett.} {\bf 83}
(1999) 3370.
\bibitem{kl} R. Kalosh, A. Linde, {\it JHEP} {\bf 02} (2000) 0005.  
\bibitem{bc} K. Behrndt, M. Cveti\'{c}, {\it Phys. Rev.} {\bf D61} (2000) 
101901. 
\bibitem{bagger} R. Altendorfer, J. Bagger, D. Nemeschansky {\it Supersymmetric Randall-Sundrum scenario}, hep-th/0003117.  
\bibitem{tgap} T. Gherghetta, A. Pomarol, {\it Bulk fields and supersymmetry 
in a slice of AdS}, hep-th/0003129.
\bibitem{falk} A. Falkowski, to be published.  
\bibitem{verlinde} H. Verlinde, {\it Holography and Compactification},
hep-th/9906182; {\it Supersymmetry at Large Distance Scales}, hep-th/0004003. 
\bibitem{ddg} K. Dienes, E. Dudas, T. Gherghetta, {\it 
Anomaly Induced Gauge Unification and Brane/Bulk Couplings in Gravity-Localized Theories},
hep-th/9908530.
 
\end{thebibliography}
\end{document}